\documentclass[envcountreset]{llncs}

\usepackage[utf8]{inputenc}
\usepackage[T1]{fontenc}
\usepackage{lmodern}
\usepackage{microtype}
\usepackage{longtable,helvet,courier,fancyhdr}
\usepackage{amsfonts}
\usepackage{alltt}
\usepackage{chemarr}
\usepackage[version=3]{mhchem}
\usepackage{amsmath,amssymb,amsfonts,bbm}
\usepackage[linesnumbered,tworuled,vlined]{algorithm2e}
\usepackage{enumerate}
\usepackage{color}
\usepackage{pgf,tikz}
\usetikzlibrary{arrows, arrows.meta, bending}

\usepackage{graphicx}

\usepackage{hyperref}
\def \Z{\mathbb{Z}}

\newcommand{\difft}[1]{\dot{#1}}

\newcommand{\phmax}{\phantom{\max\mathopen(\kern2pt}}

\newcommand{\hquad}{\kern0.5em}

\newcommand{\btwsix}{Biomod-26\xspace}
\newcommand{\btweight}{Biomod-28\xspace}

\newcommand{\bemerkung}[1]{\relax}

\begin{document}
\title{Symbolic Versus Numerical Computation and Visualization of Parameter Regions for Multistationarity of Biological Networks}
\author{
Matthew England\inst{1}\and
Hassan Errami\inst{2}\and
Dima Grigoriev\inst{3}\and
Ovidiu Radulescu\inst{4} \and
Thomas Sturm\inst{5,6} \and
Andreas Weber\inst{2}
}
\institute{
Faculty of Engineering, Environment and Computing,
  Coventry University, UK
  \email{Matthew.England@coventry.ac.uk}
  \and
Institut für Informatik II, Universität Bonn,
Germany
\email{\{errami,weber\}@cs.uni-bonn.de}
\and
CNRS, Math\'ematiques, Universit\'e de Lille, Villeneuve d'Ascq,
France
\email{Dmitry.Grigoryev@math.univ-lille1.fr}
\and
DIMNP UMR CNRS/UM 5235, University of Montpellier,
France
\email{ovidiu.radulescu@umontpellier.fr}
\and
University of Lorraine, CNRS, Inria, and LORIA, Nancy, France
\email{thomas.sturm@loria.fr}
\and
MPI Informatics and Saarland University, Saarbrücken, Germany
\email{sturm@mpi-inf.mpg.de}
}

\maketitle

\begin{abstract}
  We investigate models of the mitogenactivated protein kinases (MAPK) network,
  with the aim of determining where in parameter space there exist multiple
  positive steady states. We build on recent progress which combines various
  symbolic computation methods for mixed systems of equalities and inequalities.
  We demonstrate that those techniques benefit tremendously from a newly
  implemented graph theoretical symbolic preprocessing method. We compare
  computation times and quality of results of numerical continuation methods
  with our symbolic approach before and after the application of our
  preprocessing.
\end{abstract}


\section{Introduction}

The mathematical modelling of intra-cellular biological processes has been using
nonlinear ordinary differential equations since the early ages of mathematical
biophysics in the 1940s and 50s \cite{rashevsky1960mathematical}. A standard
modelling choice for cellular circuitry is to use chemical reactions with mass
action law kinetics, leading to polynomial differential equations. Rational
functions kinetics (for instance the Michaelis-Menten kinetics) can generally be
decomposed into several mass action steps. An important property of biological
systems is their multistationarity which means having multiple stable steady
states. Multistationarity is instrumental to cellular memory and cell
differentiation during development or regeneration of multicellular organisms
and is also used by micro-organisms in survival strategies. It is thus important
to determine the parameter values for which a biochemical model is
multistationary. With mass action reactions, testing for multiple steady states
boils down to counting real positive solutions of algebraic systems.
\newpage
  
The models benchmarked in this paper concern intracellular signaling pathways.
These pathways transmit information about the cell environment by inducing
cascades of protein modifications (phosphorylation) all the way from the plasma
membrane via the cytosol to genes in the cell nucleus. Multistationarity of
signaling usually occurs as a result of activation of upstream signaling
proteins by downstream components \cite{BhallaLyengar99}. A different mechanism
for producing multistationarity in signaling pathways was proposed by Kholodenko
\cite{Markevich2004}. In this mechanism the cause of multistationarity are
multiple phosphorylation/ dephosphorylation cycles that share enzymes. A simple,
two steps phosphorylation/dephosphorylation cycle is capable of
ultrasensitivity, a form of all or nothing response with no multiple steady
states (Goldbeter--Koshland mechanism). In multiple
phosphorylation/dephosphorylation cycles, enzyme sharing provides competitive
interactions and positive feedback that ultimately leads to multistationarity
\cite{Markevich2004,legewie2007competing}.

Our study is complementary to works applying numerical methods to ordinary
differential equations models used for biology applications. Gross et al.
\cite{gross2016numerical} used polynomial homotopy continuation methods for
global parameter estimation of mass action models. Bifurcations and
multistationarity of signaling cascades was studied with numerical methods based
on the Jacobian matrix \cite{zumsande2010bifurcations}. Other symbolic
approaches to multistationarity either propose necessary conditions or work for
particular networks
\cite{Conradi2008,ConradiMincheva,JoshiShiu2015,PerezMillan2015}.

Our work here follows \cite{Bradford2017}, where it was demonstrated that
determination of multistationarity of an 11-dimensional model of a
mitogen-activated protein kinases (MAPK) cascade can be achieved by currently
available symbolic methods when numeric values are known for all but potentially
one parameter.
We show that the symbolic methods used in \cite{Bradford2017}, viz.~real triangularization and cylindrical algebraic decomposition, and also polynomial homotopy continuation methods, benefit tremendously from a graph theoretical symbolic preprocessing method. This method has been sketched by Grigoriev et al. \cite{Grigoriev2015} and has been used for a ``hand computation,'' but had not been implemented before.
For our experiments we use the model already investigated in \cite{Bradford2017}
and a higher dimensional model of the MAPK cascade.

\section{The Systems for the Case Studies}
\label{secMAPK:system}

For our investigations we use models of the MAPK cascade that can be found in
the Biomodels database\footnote{\url{http://www.ebi.ac.uk/biomodels-main/}} as
numbers 26 and 28 \cite{Li2010a}. We refer to those models as \btwsix and
\btweight, respectively.

\subsection{\btwsix} 
\label{SEC:Sys26Def}

\btwsix, which we have studied also in \cite{Bradford2017}, is given by the
following set of differential equations. We have renamed the species names as
$x_1, \ldots, x_{11}$ and the rate constants as $k_1, \ldots, k_{16}$ to
facilitate reading:
\begin{eqnarray}
  \difft{x}_1 & = & k_{2} x_{6} + k_{15} x_{11} - k_{1} x_{1} x_{4} - k_{16} x_{1} x_{5} \nonumber \\
  \difft{x}_2 & = & k_{3} x_{6} + k_{5} x_{7} + k_{10} x_{9} + k_{13} x_{10} - x_{2} x_{5} (k_{11} + k_{12}) - k_{4} x_{2} x_{4}\nonumber \\
  \difft{x}_3 & = & k_{6} x_{7} + k_{8} x_{8} - k_{7} x_{3} x_{5}\nonumber \\
  \difft{x}_4 & = & x_{6} (k_{2} + k_{3}) + x_{7} (k_{5} + k_{6}) - k_{1} x_{1} x_{4} - k_{4} x_{2} x_{4}\nonumber \\
  \difft{x}_5 & = & k_{8} x_{8} + k_{10} x_{9} + k_{13} x_{10} + k_{15} x_{11} - x_{2} x_{5} (k_{11} + k_{12}) - k_{7} x_{3} x_{5} - k_{16} x_{1} x_{5}\nonumber \\
  \difft{x}_6 & = & k_{1} x_{1} x_{4} - x_{6} (k_{2} + k_{3})\nonumber \\
  \difft{x}_7 & = & k_{4} x_{2} x_{4} - x_{7} (k_{5} + k_{6})\nonumber \\
  \difft{x}_8 & = & k_{7} x_{3} x_{5} - x_{8} (k_{8} + k_{9})\nonumber \\
  \difft{x}_9 & = & k_{9} x_{8} - k_{10} x_{9} + k_{11} x_{2} x_{5}\nonumber \\
  \difft{x}_{10} & = &   k_{12} x_{2} x_{5} - x_{10} (k_{13} + k_{14})\nonumber \\
  \difft{x}_{11} & = &   k_{14} x_{10} - k_{15} x_{11} + k_{16} x_{1} x_{5}
\label{EQ:thesystem26}
\end{eqnarray}
The Biomodels database also gives us meaningful values for the rate constants,
which we generally substitute into the corresponding systems for our purposes
here:
\begin{align}
  k_{1} &= 0.02,&
  k_{2} &= 1,&
  k_{3} &= 0.01,&
  k_{4} &= 0.032,\nonumber\\
  k_{5} &= 1,&
  k_{6} &= 15,&
  k_{7} &= 0.045,&
  k_{8} &= 1,\nonumber\\
  k_{9} &= 0.092,&
  k_{10} &= 1,&
  k_{11} &= 0.01,&
  k_{12} &= 0.01,\nonumber\\
  k_{13} &= 1,&
  k_{14} &= 0.5,&
  k_{15} &= 0.086,&
  k_{16} &= 0.0011.\label{EQ:rcestimates26}
\end{align}
Using the left-null space of the stoichiometric matrix under positive conditions
as a conservation constraint \cite{Famili2003} we obtain three linear
conservation laws:
\begin{eqnarray}
  x_{5}  + x_{8} + x_{9} + x_{10} + x_{11} &=&  k_{17}, \nonumber \\
  x_{4}  + x_{6} + x_{7} &=&  k_{18},\nonumber \\
  x_{1}  + x_{2} + x_{3} + x_{6} + x_{7} + x_{8} + x_{9} + x_{10} + x_{11} &=&  k_{19},
\label{EQ:claws26}
\end{eqnarray}
where $k_{17}$, $k_{18}$, $k_{19}$ are new constants computed from the initial
data. Those constants are the parameters that we are interested in here.

The steady state problem for the MAPK cascade can now be formulated as a real
algebraic problem as follows. We replace the left hand sides of all equations in
(\ref{EQ:thesystem26}) with $0$ and substitute the values from
(\ref{EQ:rcestimates26}). This together with (\ref{EQ:claws26}) yields a system
of parametric polynomial equations with polynomials in
$\Z[k_{17},k_{18},k_{19}][x_1,\dots,x_{11}]$. Since all entities in our model
are strictly positive, we add to our system positivity conditions $k_{17}>0$,
$k_{18}>0$, $k_{19}>0$ and $x_1>0$, \dots,~$x_{11}>0$. In terms of first-order
logic the conjunction over our equations and inequalities yields a
quantifier-free Tarski formula.

\subsection{\btweight}

The system with number 28 in the Biomodels database
is given by the following set of differential equations. Again,
we have renamed the species names into $x_1, \ldots, x_{16}$ and the rate constants into
$k_1, \ldots, k_{27}$ to facilitate reading:

\begin{eqnarray*}
\difft{x}_{1} & = & k_2 x_9 + k_8 x_{10} + k_{21} x_{15} + k_{26} x_{16} - k_1 x_1 x_5 - k_7 x_1 x_5 - k_{22} x_1 x_6 - k_{27} x_1 x_6  \nonumber \\
\difft{x}_{2} &= & k_3 x_9 + k_5 x_7 + k_{24} x_{12} - k_4 x_2 x_5 - k_{23} x_2 x_6 \nonumber \\
\difft{x}_{3} & = & k_9 x_{10} + k_{11} x_8 + k_{16} x_{13} + k_{19} x_{14} - k_{10} x_3 x_5 - k_{17} x_3 x_6 - k_{18} x_3 x_6 \nonumber \\
\difft{x}_{4} &=&  k_6 x_7 + k_{12} x_8 + k_{14} x_{11} - k_{13} x_4 x_6 \nonumber \\
\difft{x}_{5} &= & k_2 x_9 + k_3 x_9 + k_5 x_7 + k_6 x_7 + k_8 x_{10} + k_9 x_{10} + k_{11} x_8 + k_{12} x_8 -\nonumber \\
                  &    & \quad k_1 x_1 x_5 - k_4 x_2 x_5 - k_7 x_1 x_5 - k_{10} x_3 x_5  \nonumber \\
\difft{x}_{6} & = & k_{14} x_{11} + k_{16} x_{13} + k_{19} x_{14} + k_{21} x_{15} + k_{24} x_{12} + k_{26} x_{16} - \nonumber \\
                &    & \quad  k_{13} x_4 x_6 - k_{17} x_3 x_6 - k_{18} x_3 x_6 - k_{22} x_1 x_6 - k_{23} x_2 x_6 - k_{27} x_1 x_6  \nonumber \\
\difft{x}_{7} & = & k_4 x_2 x_5 - k_6 x_7 - k_5 x_7  \nonumber \\
\difft{x}_{8} &=  & k_{10} x_3 x_5 - k_{12} x_8 - k_{11} x_8  \nonumber \\
\difft{x}_{9} &= & k_1 x_1 x_5 - k_3 x_9 - k_2 x_9  \nonumber \\
\difft{x}_{10} &= &k_7 x_1 x_5 - k_9 x_{10} - k_8 x_{10}  \nonumber \\
\difft{x}_{11} &=& k_{13} x_4 x_6 - k_{15} x_{11} - k_{14} x_{11}  \nonumber \\
\difft{x}_{12}& = &  k_{23} x_2 x_6 - k_{25} x_{12} - k_{24} x_{12}  \nonumber \\
\difft{x}_{13} &= &k_{15} x_{11} - k_{16} x_{13} + k_{17} x_3 x_6  \nonumber \\
\difft{x}_{14}& = & k_{18} x_3 x_6 - k_{20} x_{14} - k_{19} x_{14}  \nonumber \\
\difft{x}_{15}& = &k_{20} x_{14} - k_{21} x_{15} + k_{22} x_1 x_6  \nonumber \\
\difft{x}_{16}& = &k_{25} x_{12} - k_{26} x_{16} + k_{27} x_1 x_6  \label{EQ:thesystem28} 
\end{eqnarray*}
The estimates of the rate constants given in the  Biomodels database are:
\begin{align*}
  k_{1} &= 0.005,&
  k_{2} &= 1,&
  k_{3} &= 1.08,&
  k_{4} &= 0.025,\nonumber\\
  k_{5} &= 1,&
  k_{6} &= 0.007,&
  k_{7} &= 0.05,&
  k_{8} &= 1,\nonumber\\
  k_{9} &= 0.008,&
  k_{10} &= 0.005,&
  k_{11} &= 1,&
  k_{12} &= 0.45,\nonumber\\
  k_{13} &= 0.045,&
  k_{14} &= 1,&
  k_{15} &= 0.092,&
  k_{16} &= 1,&\nonumber\\
  k_{17} &= 0.01,&
  k_{18} &= 0.01,&
  k_{19} &= 1,&
  k_{20} &= 0.5,&\nonumber\\
  k_{21} &= 0.086,&
  k_{22} &= 0.0011,&
  k_{23} &= 0.01,&
  k_{24} &= 1,&\nonumber\\
  k_{25} &= 0.47,&
  k_{26} &= 0.14,&
  k_{27} &= 0.0018.  \label{EQ:rcestimates28}
\end{align*}
Again, using the left-null space of the stoichiometric matrix under positive conditions as a conservation constraint \cite{Famili2003}
we obtain the following:
\begin{eqnarray*}
  x_6 + x_{11} + x_{12} + x_{13}+ x_{14} + x_{15} + x_{16}   &=&  k_{28}, \nonumber \\
 x_5 + x_7 + x_8 + x_9 + x_{10} &=& k_{29} ,\nonumber \\
 x_1 + x_2 + x_3 + x_4 + x_7 + x_8 + x_9 + x_{10} + x_{11} + {} & &\nonumber \\
 \quad  x_{12} + x_{13} + x_{14} + x_{15} + x_{16}  &=& k_{30}, 
\label{EQ:claws28}
\end{eqnarray*}
where $k_{28}$, $k_{29}$, $k_{30}$ are new constants computed from the initial data.
We formulate the real algebraic problem as described at the end of
Sect.~\ref{SEC:Sys26Def}. In particular, note that we need positivity conditions
for all variables and parameters.

\section{Graph-Theoretical Symbolic Preprocessing}

The complexity, primarily in terms of dimension, of polynomial systems obtained
with steady-state approximations of biological models plus conservation laws is
comparatively high for the application of symbolic methods. It is therefore
highly relevant for the success of such methods to identify and exploit
particular structural properties of the input. Our models have remarkably low
total degrees with many linear monomials after some substitutions for rate
constants. This suggests to preprocess with essentially Gaussian elimination in
the sense of solving single suitable equations with respect to some variable and
substituting the corresponding solution into the system.

Generalizing this idea to situations where linear variables have parametric
coefficients in the other variables requires, in general, a parametric variant
of Gaussian elimination, which replaces the input system with a finite case
distinction with respect to the vanishing of certain coefficients and one
reduced system for each case. With \btwsix and \btweight considered here it
turns out that the positivity assumptions on the variables are strong enough to
effectively guarantee the non-vanishing of all relevant coefficients so that
case distinctions are never necessary.
On the other hand, those positivity conditions establish an apparent obstacle,
because we are formally not dealing with a parametric system of linear equations
but with a parametric linear programming problem. However, here the theory of
real quantifier elimination by virtual substitution tells us that it is
sufficient that the inequality constraints play a passive role. Those
constraints must be considered when substituting Gauss solutions from the
equations, but otherwise can be ignored \cite{LoosWeispfenning:93a,Kosta:16a}.

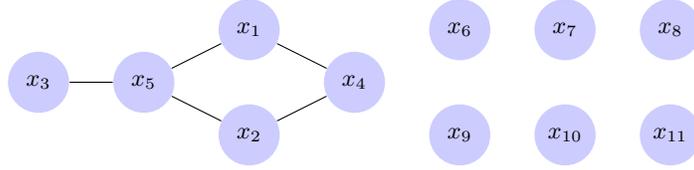
\begin{figure}[t]
  \centering
  \begin{tikzpicture}[scale=.7,auto=left,every
    node/.style={circle,fill=blue!20,minimum size = 2.5em}]
    \node (n3) at (1,20) {$x_3$}; \node (n5) at (3,20) {$x_5$}; \node (n1) at
    (5,21) {$x_1$}; \node (n2) at (5,19) {$x_2$}; \node (n4) at (7,20)
    {$x_4$};

    \node (n6) at (9,21) {$x_6$}; \node (n7) at (11,21) {$x_7$}; \node (n8) at
    (13,21) {$x_8$}; \node (n9) at (9,19) {$x_9$}; \node (n10) at (11,19)
    {$x_{10}$}; \node (n11) at (13,19) {$x_{11}$};

    \foreach \from/\to in {n3/n5,n1/n5,n1/n4,n2/n4,n2/n5} \draw (\from) --
    (\to);
  \end{tikzpicture}
  \caption{The graph for \btwsix is loosely connected. Its minimum vertex
    cover $\{x_4,x_5\}$ is small. All other variables form a maximum independent
    set, which can be eliminated with linear methods.\label{fig:vc}}
\end{figure}

Parametric Gaussian elimination can increase the degrees of variables in the
parametric coefficient, in particular destroying their linearity and suitability
to be used for further reductions. As an example consider the steady-state
approximation, i.e., all left hand sides replaced with $0$, of the system in
(\ref{EQ:thesystem26}), solving the last equation for $x_5$, and substituting
into the first equation. The natural question for an optimal strategy to
Gauss-eliminate a maximal number of variables has been answered positively only
recently~\cite{Grigoriev2015}: draw a graph, where vertices are variables and
edges indicate multiplication between variables within some monomial. Then one
can Gauss-eliminate a \emph{maximum independent set}, which is the complement of
a \emph{minimum vertex cover}. Fig.~\ref{fig:vc} shows that graph for \btwsix,
where $\{x_4,x_5\}$ is a minimal vertex cover, and all other variables can be
linearly eliminated. Similarly, for \btweight we find $\{x_5,x_6\}$ as a minimum
vertex cover. Recall that minimum vertex cover is one of Karp's 21 classical NP
complete problems \cite{Karp:72}. However, our instances considered here and
instances to be expected from other biological models are so small that the use
of existing approximation algorithms \cite{Grandoni2008} appears unnecessary. We
have used real quantifier elimination, which did not consume measurable CPU
time; alternatively one could use integer linear programming or SAT-solving.

It is a most remarkable fact that a significant number of biological models in
the databases have that property of loosely connected variables. This phenomenon
resembles the well-known \emph{community structure} of propositional
satisfiability problems, which has been identified as one of the key structural
reasons for the impressive success of state-of-the-art CDCL-based SAT solvers
\cite{girvan2002community}.

We conclude this section with the reduced systems as computed with our
implementation in Redlog~\cite{DolzmannSturm:97a}. For \btwsix we obtain
$x_{5} >0$, $x_{4} > 0$, $k_{19} > 0$, $k_{18}> 0$, $k_{17} > 0$ and
{\small\begin{eqnarray*}
  1062444 k_{18} x_{4}^{2} x_{5} + 23478000 k_{18} x_{4}^{2} + 1153450 k_{18}
  x_{4} x_{5}^{2} + 2967000 k_{18} x_{4} x_{5} &&\\
  {} + 638825 k_{18} x_{5}^{3} + 49944500 k_{18} x_{5}^{2} - 5934 k_{19}
  x_{4}^{2} x_{5} - 989000 k_{19} x_{4} x_{5}^{2}&&\\
  {} - 1062444 x_{4}^{3} x_{5} -
  23478000 x_{4}^{3} - 1153450
  x_{4}^{2} x_{5}^{2}- 2967000 x_{4}^{2} x_{5}&&\\
  {} - 638825 x_{4} x_{5}^{3} -
  49944500 x_{4} x_{5}^{2} &=& 0,\\
  1062444 k_{17} x_{4}^{2} x_{5} + 23478000 k_{17} x_{4}^{2} + 1153450
  k_{17}x_{4} x_{5}^{2} + 2967000 k_{17} x_{4} x_{5}&&\\
  {}+ 638825 k_{17}
  x_{5}^{3}
  + 49944500 k_{17} x_{5}^{2} - 1056510 k_{19} x_{4}^{2} x_{5} - 164450 k_{19}
  x_{4} x_{5}^{2}&&\\
  {}- 638825 k_{19} x_{5}^{3} - 1062444 x_{4}^{2} x_{5}^{2} - 23478000 x_{4}^{2}
  x_{5} - 1153450 x_{4} x_{5}^{3}&&\\
  {} - 2967000 x_{4} x_{5}^{2} -
  638825 x_{5}^{4} - 49944500 x_{5}^{3} &=& 0.
\end{eqnarray*}}
For \btweight we obtain
$x_{6} >0$, $x_{5} > 0$, $k_{30} > 0$, $k_{29}> 0$, $k_{28} > 0$ and
{\small\begin{eqnarray*}
  3796549898085 k_{29} x_{5}^{3} x_{6} + 71063292573000 k_{29} x_{5}^{3} +
  106615407090630 k_{29} x_{5}^{2} x_{6}^{2}&&\\
  {}+ 479383905861000 k_{29} x_{5}^{2} x_{6} + 299076127852260 k_{29} x_{5}
  x_{6}^{3}&&\\
  {}+ 3505609439955600 k_{29} x_{5} x_{6}^{2}
   + 91244417457024 k_{29} x_{6}^{4}&&\\
  {}+ 3557586742819200 k_{29} 
  x_{6}^{3} - 598701732300 k_{30} x_{5}^{3} x_{6}&&\\
  {} - 83232870778950 k_{30} x_{5}^{2} x_{6}^{2}
  - 185019487578700 k_{30} x_{5}x_{6}^{3}&&\\
   - 3796549898085 x_{5}^{4} x_{6}
   - 71063292573000 x_{5}^{4}
  - 106615407090630 x_{5}^{3}
  x_{6}^{2}&&\\
  {} - 479383905861000 x_{5}^{3} x_{6} - 299076127852260 x_{5}^{2}
  x_{6}^{3}
  - 3505609439955600 x_{5}^{2} x_{6}^{2}&&\\
  {}- 91244417457024 x_{5}
  x_{6}^{4} - 3557586742819200 x_{5} x_{6}^{3} &=& 0,
\\
  3796549898085 k_{28} x_{5}^{3} x_{6} + 71063292573000 k_{28} x_{5}^{3} +
  106615407090630 k_{28} x_{5}^{2} x_{6}^{2}&&\\
  {}+ 479383905861000 k_{28} x_{5}^{2}
  x_{6} + 299076127852260 k_{28} x_{5} x_{6}^{3}&&\\
  {}+ 3505609439955600 k_{28} x_{5} x_{6}^{2}
  + 91244417457024 k_{28} x_{6}^{4}&&\\
  {}+ 3557586742819200 k_{28} x_{6}^{3}
  - 3197848165785 k_{30} x_{5}^{3} x_{6}&&\\
  {} - 23382536311680 k_{30}
  x_{5}^{2} x_{6}^{2} - 114056640273560 k_{30} x_{5} x_{6}^{3}&&\\
  {}- 91244417457024
  k_{30} x_{6}^{4}
  - 3796549898085 x_{5}^{3} x_{6}^{2} - 71063292573000
  x_{5}^{3} x_{6}&&\\
  {}- 106615407090630 x_{5}^{2} x_{6}^{3}
  - 479383905861000
  x_{5}^{2} x_{6}^{2} - 299076127852260 x_{5} x_{6}^{4}&&\\
  {} - 3505609439955600 x_{5}
  x_{6}^{3} - 91244417457024 x_{6}^{5} - 3557586742819200 x_{6}^{4} &=& 0.
\end{eqnarray*}}%
Notice that no complex positivity constraints come into existence with these
examples. All corresponding substitution results are entailed by the other
constraints, which is implicitly discovered by using the standard simplifier
from \cite{DolzmannSturm:97c} during preprocessing.

\section{Determination of Multiple Steady States}
\label{SEC:Grid}


We aim to identify via grid sampling regions of parameter space where
multistationarity occurs. Our focus is on the identification of regions with
multiple positive real solutions for the parameters introduced with the
conservation laws. We will encounter one or three such solutions and allow
ourselves for biological reasons to assume monostability or bistability,
respectively. Furthermore, a change in the number of solutions between one and
three is indicative of a saddle-node bifurcation between a monostable and a
bistable case. A mathematically rigorous treatment of stability would, possibly
symbolically, analyze the eigenvalues of the Jacobian of the respective
polynomial vector field. We consider two different approaches: first a
polynomial homotopy continuation method implemented in Bertini, and second a
combination of symbolic computation methods implemented in Maple. We compare the
approaches with respect to performance and quality of results for both the
reduced and the unreduced systems.

\subsection{Numerical Approach}
\label{SEC:Bertini}

We use the homotopy solver Bertini \cite{BHSW06} in its standard configuration
to compute complex roots. We parse the output of Bertini using Python, and
determined numerically, which of the complex roots are real and positive using a
threshold of $10^{-6}$ for positivity. Computations are done in Python with
Bertini embedded.

For System \btwsix we produced the two plots in
Fig.~\ref{FIG:Bertini-Sys26-Original} using the original system and the two in
Fig.~\ref{FIG:Bertini-Sys26-Reduced} using the reduced system. The sampling
range for $k_{19}$ was from 200 to 1000 by 50. In the left plots the sampling
range for $k_{17}$ is from 80 to 200 by 10 with $k_{18}$ fixed at 50. In the
right plots the sampling range for $k_{18}$ is 5 to 75 by 5 with $k_{17}$ fixed
to 100. We see two regions forming according to the number of fixed points: yellow discs indicate one fixed point and blue boxes three.  The diamonds indicate numerical errors where zero (red) or two (green) fixed states were identified.    We analyse these further in Sect.~\ref{SEC:comp}.

\begin{figure}[p]
  \setlength{\abovecaptionskip}{5pt}
  \setlength{\belowcaptionskip}{10pt plus 5pt}
  \centering
  \includegraphics[width=0.44\textwidth]{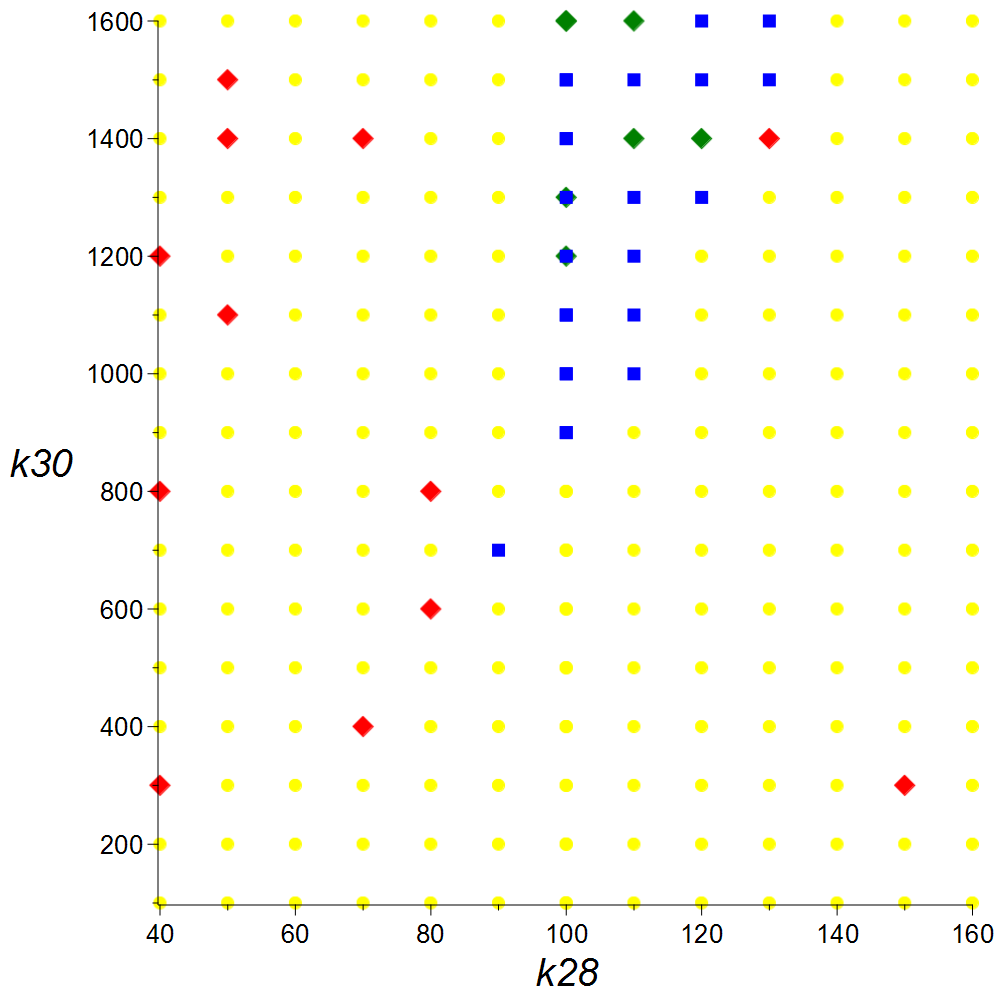}
  \includegraphics[width=0.44\textwidth]{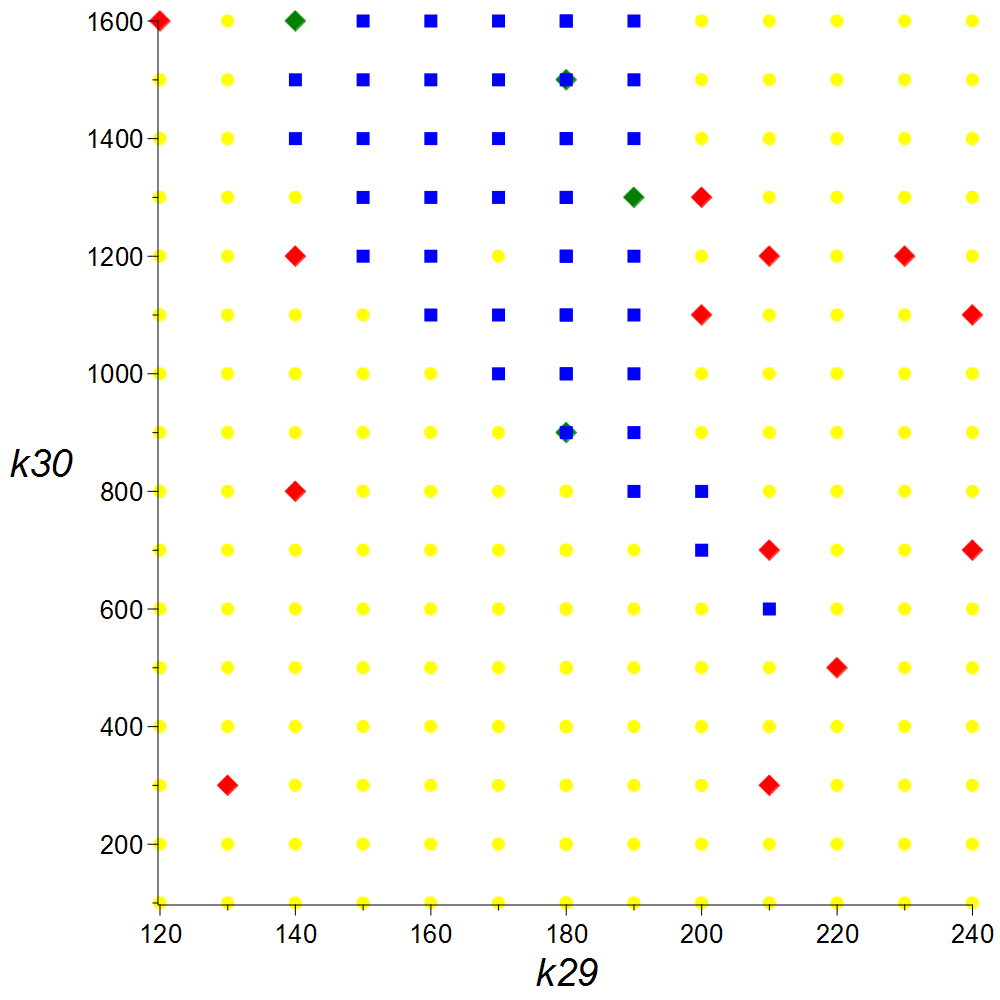}
  \caption{Bertini grid sampling on the original version of \btweight (see
    Sect.~\ref{SEC:Bertini})\label{FIG:Bertini-Sys28-Original}}
\end{figure}

For \btweight we produced the two plots in Fig.~\ref{FIG:Bertini-Sys28-Original}
using the original system.
The sampling range for $k_{30}$ was from 100 to 1600 by 100. In the left plots
the sampling range for $k_{28}$ is from 40 to 160 by 10 with $k_{29}$ fixed at
180. In the right plots the sampling range for $k_{29}$ is from 120 to 240 by 10
with $k_{28}$ fixed to 100. The colours and shapes indicate the number of fixed points as
before. For the reduced system Bertini (wrongly) could not find any roots (not even complex
ones) for any of the parameter settings. The situation did not change when
going from adaptive precision to a very high fixed precision. However, we have
not attempted more sophisticated techniques like providing user homotopies. We
analyse these results further in Sect.~\ref{SEC:comp}.

\subsection{Symbolic Approach}
\label{SEC:Maple}

Our next approach will still use grid sampling, but each sample point will
undergo a symbolic computation. The result will still be an approximate
identification of the region (since the sampling will be finite) but the results
at those sample points will be guaranteed free of numerical errors. The
computations follow the strategy introduced in \cite[Section 2.1.2]{Bradford2017}.
This combined tools from the Regular Chains
Library\footnote{\url{http://www.regularchains.org/}} available for use in Maple. Regular
chains are the triangular decompositions of systems of polynomial equations
(triangular in terms of the variables in each polynomial). Highly efficient
methods for working in complex space have been developed based on these (see
\cite{Wang2000} 
for a survey).

We make use of recent work by Chen et al.~\cite{CDMMXX13} which adapts these
tools to the real analogue: semi-algebraic systems. They describe algorithms to
decompose any real polynomial system into finitely many regular semi-algebraic
systems: both directly and by computation of components by dimension. The latter
(the so called \emph{lazy} variant) was key to solving the 1-parameter MAPK
problem in \cite{Bradford2017}. However, for the zero dimensional computations
of this paper there is only one solution component and so no savings from lazy
computations.

For a given system and sample point we apply the real triangularization (RT) on the quantifier-free formula (as described at the end of Sect.~\ref{SEC:Sys26Def}: a quantifier free conjunction of equities and inequalities) evaluated with the parameter estimates and sample point values.  
This produces a simplified system in several senses.  
First, as guaranteed by the algorithm, the output is triangular according to a variable ordering.  
So there is a univariate component, then a bivariate component introducing one more variable and so on.   
Secondly, for all the MAPK models we have studied so far, all but the final (univariate) of these equations has been linear in its main variable.  
This thus allows for easy back substitution.  Thirdly, most of the positivity conditions are implied by the output rather than being an explicit part of it, 
in which case a simpler sub-system can be solved and back substitution performed instantly.  

\subsubsection{\btwsix}

For the original version of \btwsix the output of RT was a component consisting
of 11 equations and a single inequality. The equations were in ascending main
variable according to the provided ordering (same as the labelling). All but the
final equation is linear in its main variable, with the final equation being
univariate and degree 6 in $x_1$. The output of the triangularization requires
that this variable be positive, $x_1>0$, with the positivity of the other
variables implied by solutions to the system.  So to proceed we must find the positive real roots of the degree 8 univariate polynomial in $x_1$: counting these will imply the number of real positive solutions of the parent system. We do this using the root isolation tools in the Regular Chains Library.  This whole process was performed iteratively for the same sampling regime as Bertini used to produce Fig.~\ref{FIG:Maple-Sys26}.

We repeated the process on the reduced version of the system. The
triangularization again reduced the problem to univariate real root isolation,
this time with only one back substitution step needed. As to be expected from a
fully symbolic computation, the output is identical and so again represented by
Fig.~\ref{FIG:Maple-Sys26}. However, the computation was significantly 
quicker with this reduced system. More details are given in the comparison in
Sect.~\ref{SEC:comp}.

\subsubsection{\btweight}

\begin{figure}[p]
  \setlength{\abovecaptionskip}{5pt}
  \setlength{\belowcaptionskip}{10pt plus 5pt}
  \centering
  \includegraphics[width=0.44\textwidth]{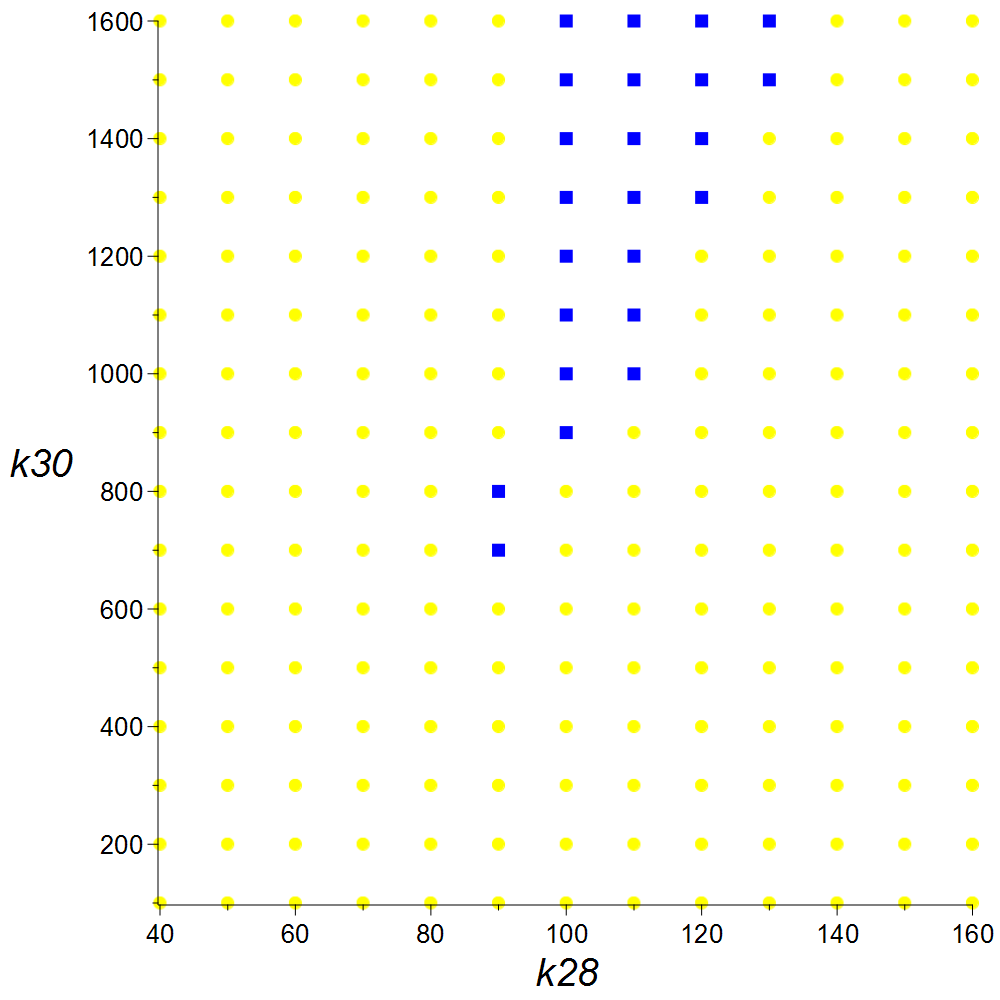}
  \includegraphics[width=0.44\textwidth]{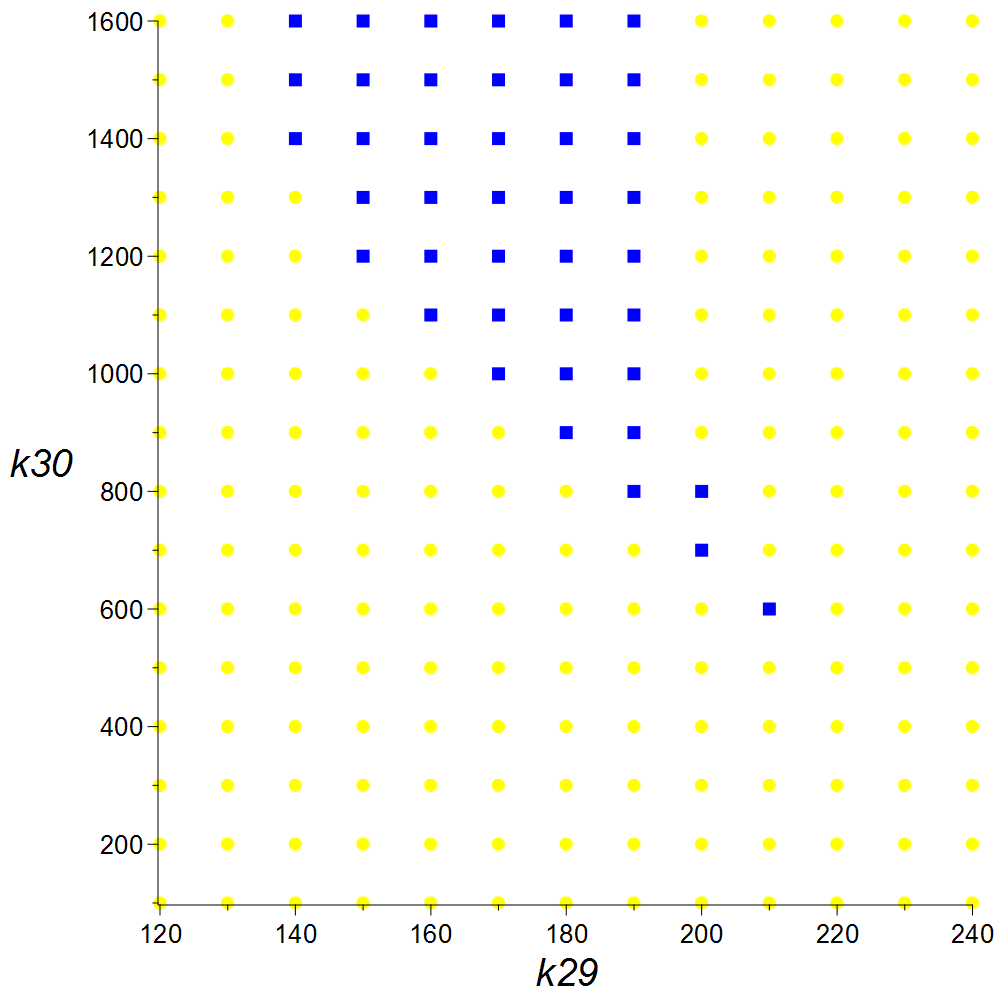}
  \caption{Maple grid sampling on \btweight (see Sect.~\ref{SEC:Maple})\label{FIG:Maple-Sys28}}
\end{figure}

The same process was conducted on \btweight. As with \btwsix the system was
triangular with all but the final equation linear in its main variable; this
time the final equation is degree 8. However, unlike \btwsix two positivity
conditions were returned in the output meaning we must solve a bivariate problem
before we can back substitute to the full system. Rather than just perform
univariate real root isolation we must build a Cylindrical Algebraic
Decomposition (CAD) (see, e.g., \cite{BDEMW16} and the references within) sign
invariant for the final two equations and interrogate its cells to find those
where the equations are satisfied and variable positive. Counting these we find
always 1 or 3 cells, with the latter indicating bistability. This is similar to
the approach used in \cite{Bradford2017}, although in that case the 2D CAD was
for one variable and one parameter. We used the implementation of CAD in the
Regular Chains Library \cite{CMXY09,BCDEMW14} with the results producing the plots in Fig.~\ref{FIG:Maple-Sys28}.

For the reduced system we proceeded similarly. A 2D CAD still needed to be
produced after triangularization and so in this case there was no reduction in the number of equations to study with CAD via back substitution. 
However, it was still beneficial to pre-process CAD with real triangularization: the average time per sample point with pre-processing (and including time taken to pre-process) was 0.485 seconds while without it was 3.577 seconds.

\subsection{Comparison}
\label{SEC:comp}



\begin{figure}[p]
  \setlength{\abovecaptionskip}{5pt}
  \setlength{\belowcaptionskip}{10pt plus 5pt}
  \centering
  \includegraphics[width=0.44\textwidth]{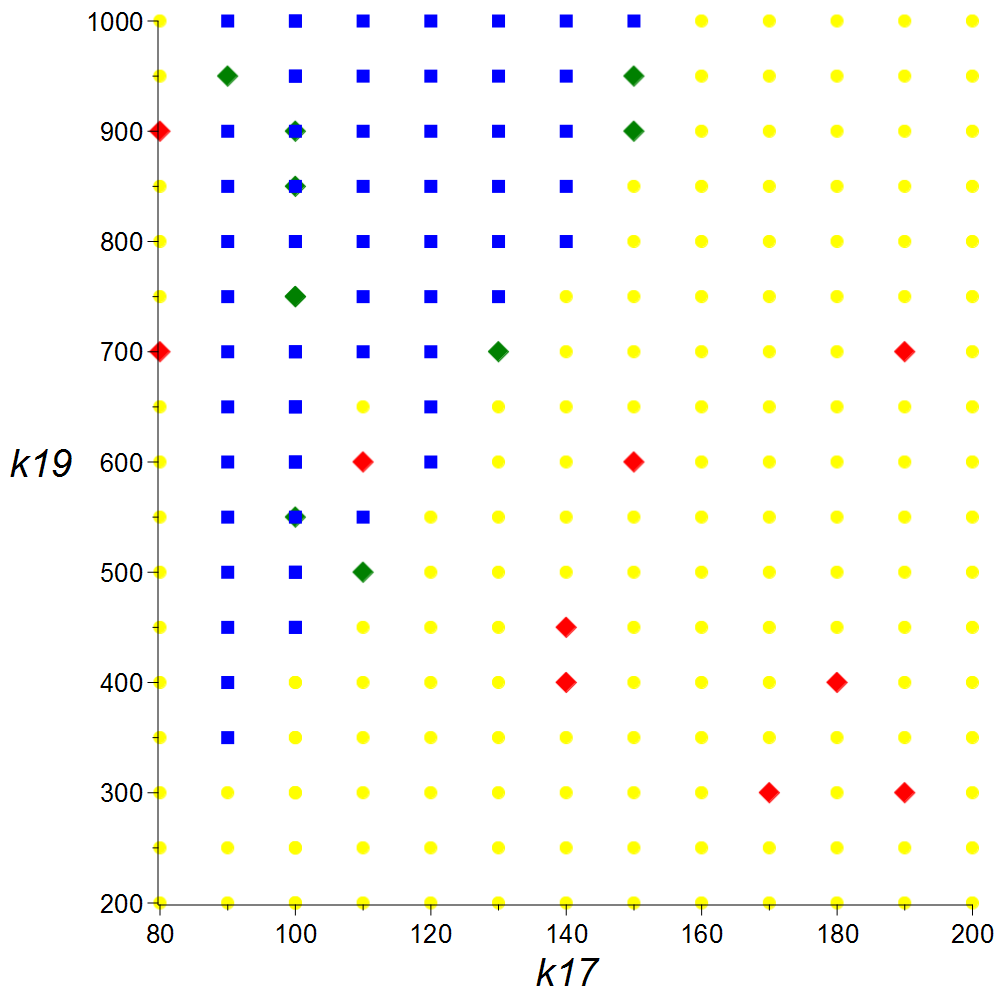}
  \includegraphics[width=0.44\textwidth]{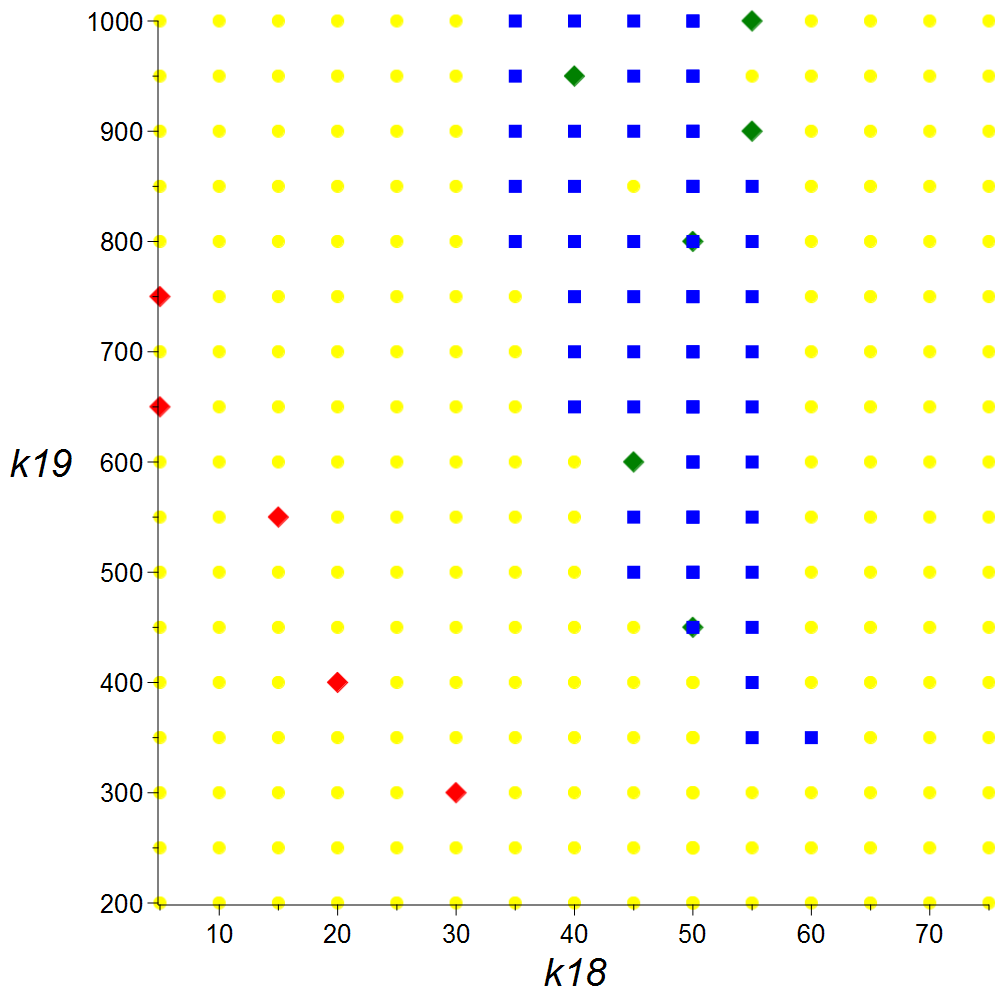}
  \caption{Bertini grid sampling on the original version of \btwsix (see
    Sect.~\ref{SEC:Bertini})\label{FIG:Bertini-Sys26-Original}}
\end{figure} 

\begin{figure}[p]
  \setlength{\abovecaptionskip}{5pt}
  \setlength{\belowcaptionskip}{10pt plus 5pt}
  \centering
  \includegraphics[width=0.44\textwidth]{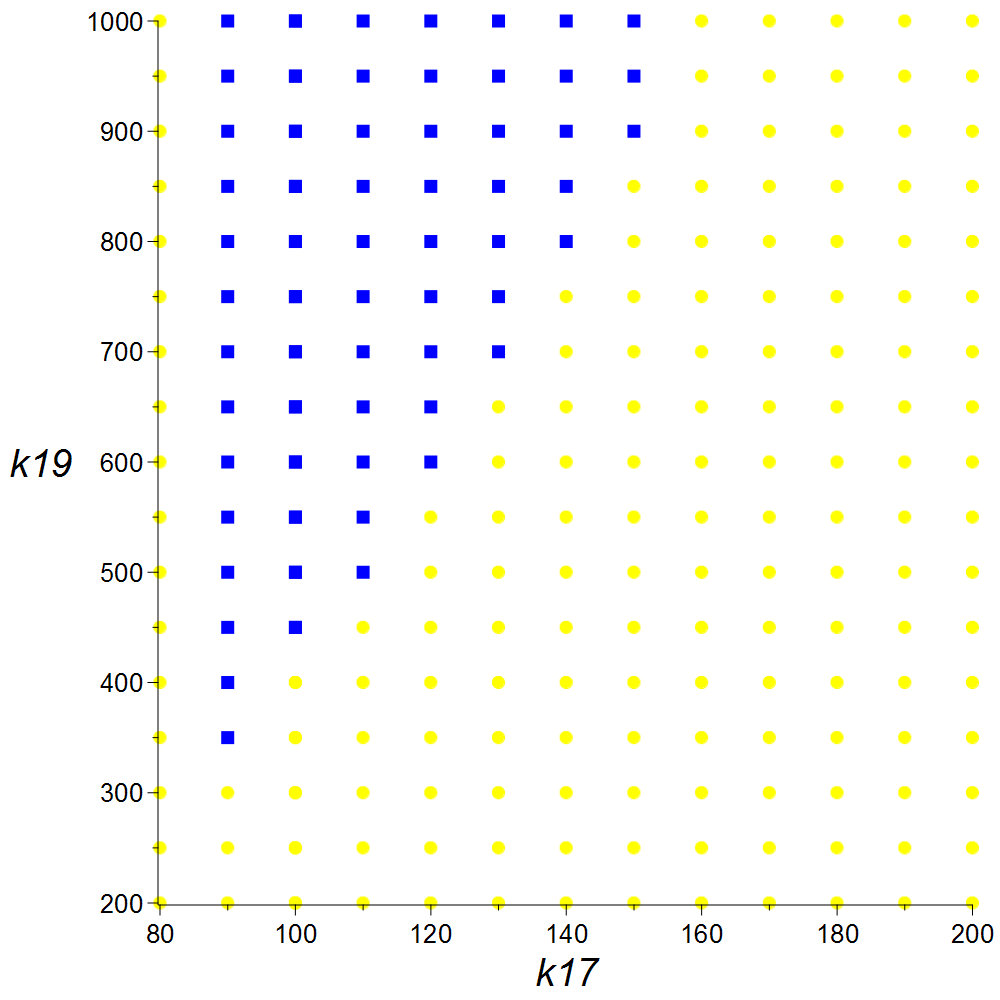}
  \includegraphics[width=0.44\textwidth]{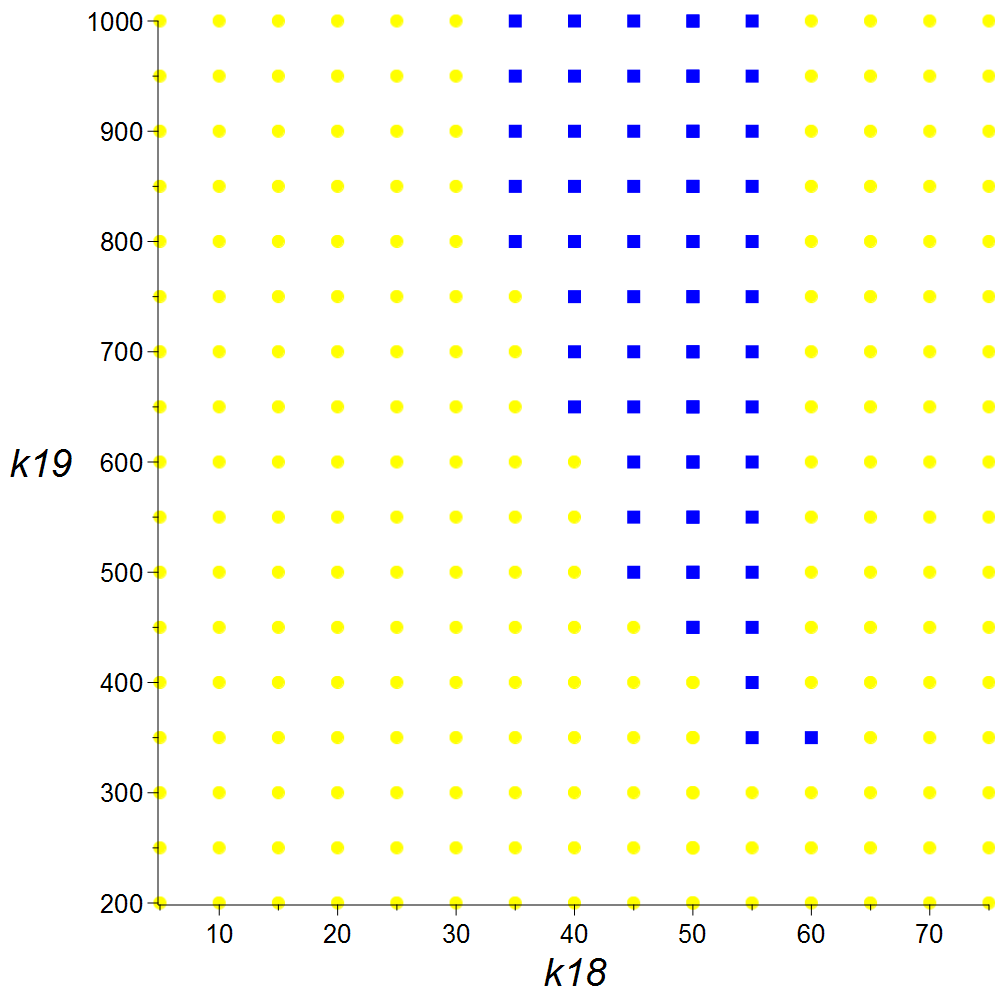}
  \caption{Bertini grid sampling on the reduced version of \btwsix (see
    Sect.~\ref{SEC:Bertini})\label{FIG:Bertini-Sys26-Reduced}}
\end{figure} 

\begin{figure}[p]
  \setlength{\abovecaptionskip}{5pt}
  \setlength{\belowcaptionskip}{0pt}
  \centering
  \includegraphics[width=0.44\textwidth]{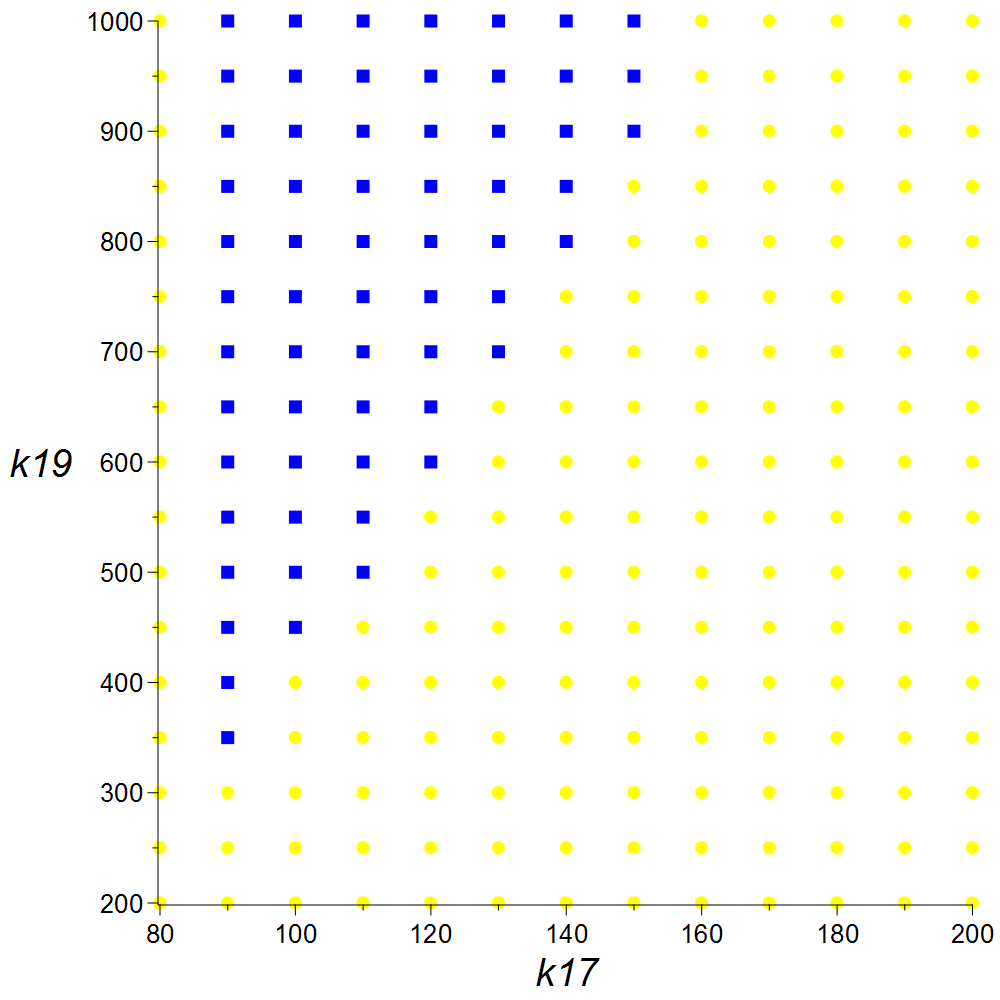}
  \includegraphics[width=0.44\textwidth]{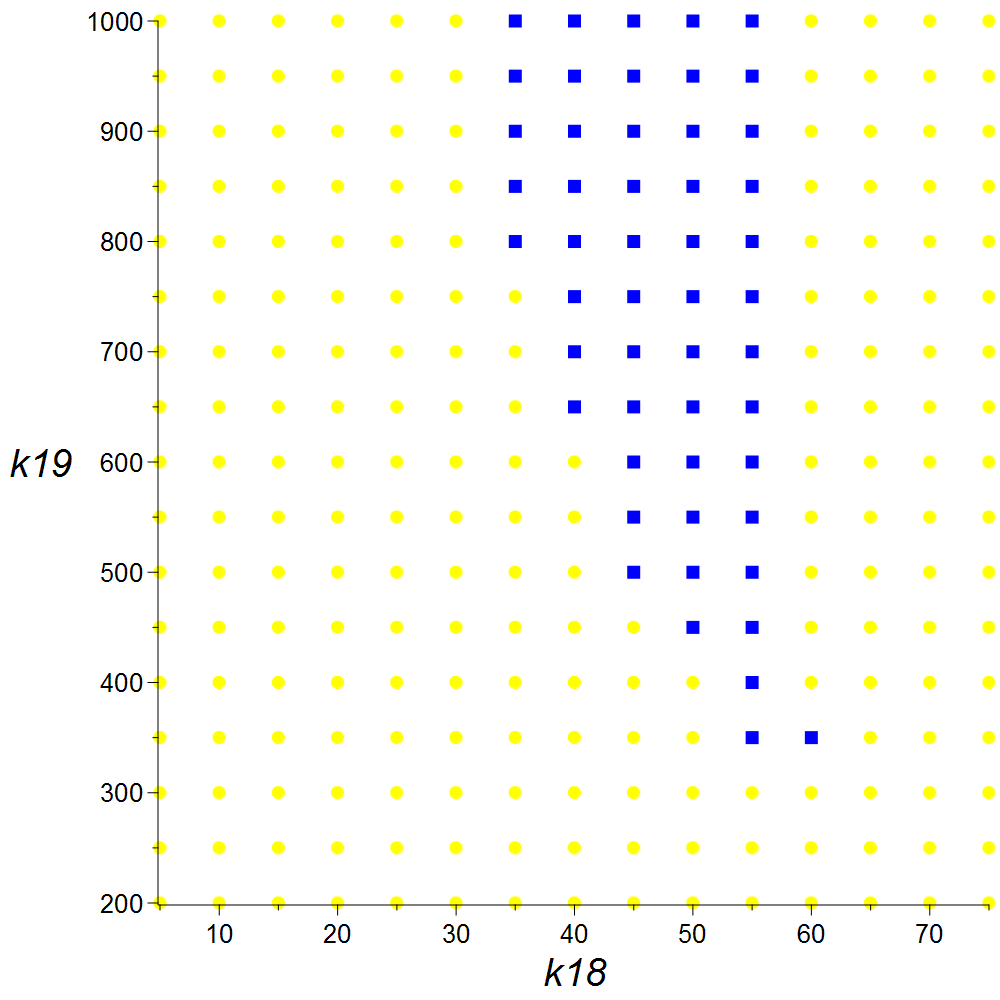}
  \caption{Maple grid sampling on \btwsix (see Sect.~\ref{SEC:Maple})\label{FIG:Maple-Sys26}}
\end{figure}




\begin{figure}[p]
  \setlength{\abovecaptionskip}{5pt}
  \setlength{\belowcaptionskip}{0pt}
  \centering
  \includegraphics[width=0.44\textwidth]{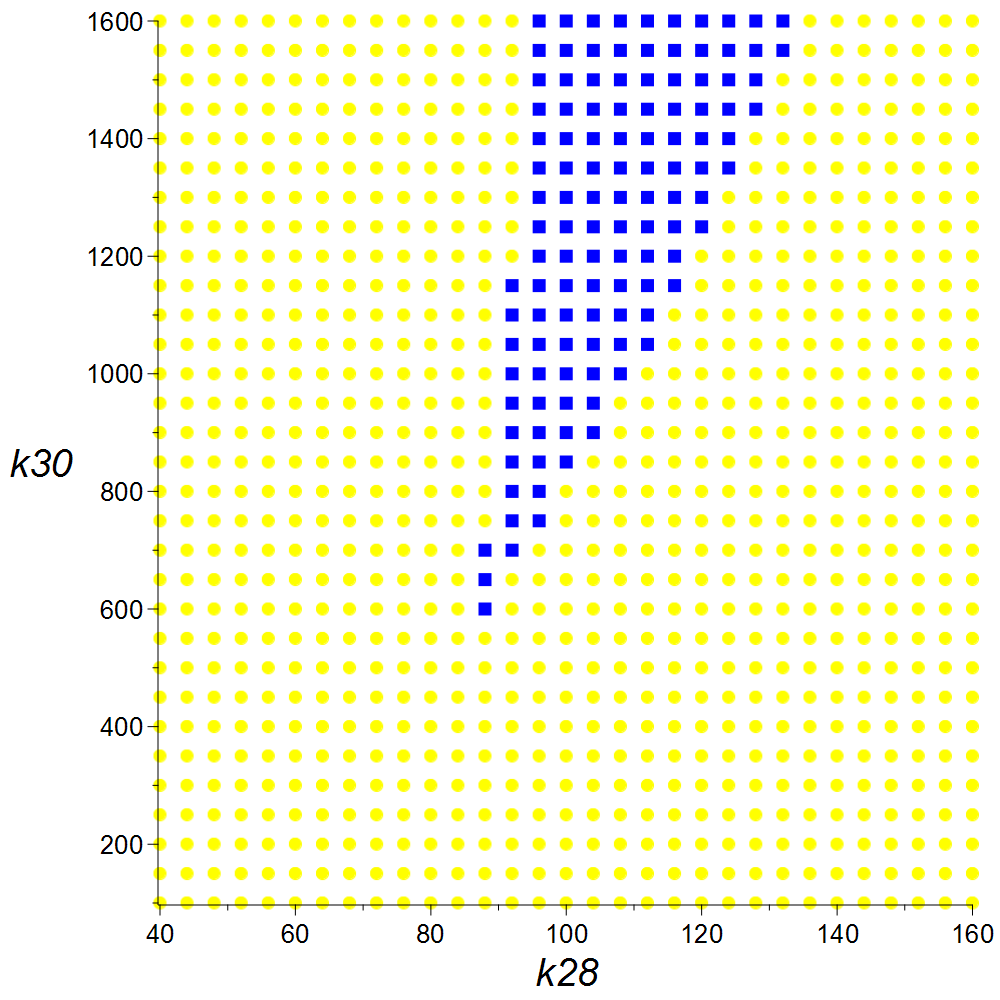}
  \includegraphics[width=0.44\textwidth]{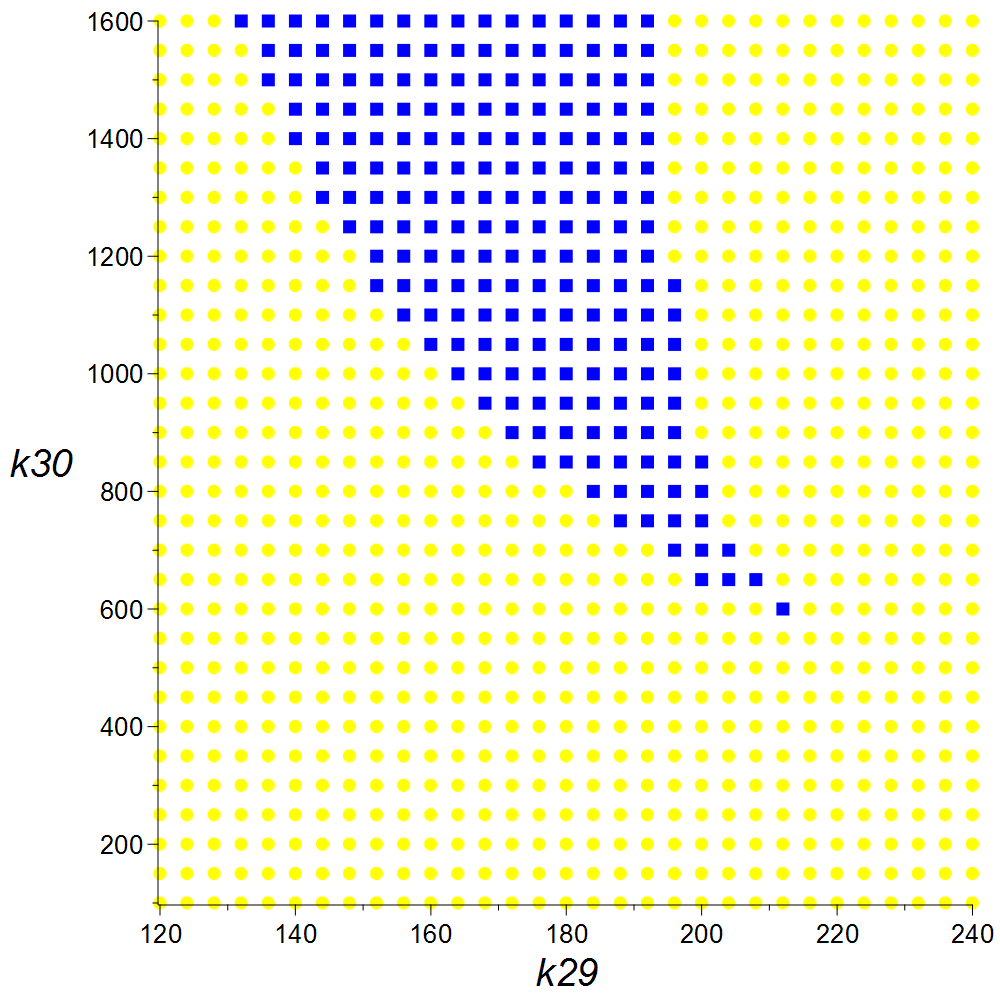}
  \caption{As Fig.~\ref{FIG:Maple-Sys28} but with a higher sampling rate\label{FIG:Sys28Detailed}}
\end{figure} 


Figure~\ref{FIG:Bertini-Sys26-Original}, Fig.~\ref{FIG:Bertini-Sys26-Reduced}, and Fig.~\ref{FIG:Maple-Sys26} all refer to \btwsix.  The latter, produced using the symbolic techniques in Maple, is guaranteed free of numerical error.  We see that computing with the reduced system rather than the original system allowed Bertini to avoid such errors: the rouge red and green diamonds in Fig.~\ref{FIG:Bertini-Sys26-Original}.  
However, in the case of \btweight the reduction led to catastrophic effects for Bertini: built-in heuristics quickly (and wrongly) concluded that
there are no zero dimensional  solutions for the system, and when switching to a positive dimensional run also no solutions could be found.

Bertini computations (v1.5.1) were carried out on a Linux  64 bit Desktop PC
with Intel i7. Maple computations (v2016 with April 2017 Regular Chains)
were carried out on a Windows 7 64 bit Desktop PC with Intel i5.  

For \btwsix the pairs of plots together contain 476 sample points.
Table~\ref{TAB:SysTime} shows timing data.  We see that both Bertini and Maple
benefited from the reduced system: Bertini took a third of the original time
while the speedup for Maple was even greater: a tenth of the original.  Also,
perhaps surprisingly, the symbolic methods were quicker than the numerical ones
here.  For \btweight the speed-up enjoyed by the symbolic methods was even
greater (almost 100 fold).  However, for this system Bertini was significantly
faster.  The symbolic methods used are well known for their doubly exponential
computational complexity (in the number of variables) so it is not
 surprising that as the system size increases there so should the results of the comparison.
We see some other statistical data for the timings in Maple: the standard deviation for the timings is fairly modest but in each row we see there are outliers many multiples of the mean value and so the median is always a little less than the mean average.


\begin{table}[t]
  \addtolength{\tabcolsep}{0.3em}
  \centering
  \caption{Timing data (in seconds) of the grid samplings described in
    Sect.~\ref{SEC:Grid}. Numerical computation is using Bertini; Symbolic
    computation is using Maple Regular Chains\label{TAB:SysTime}}
  \begin{tabular}{lr@{\qquad}rrrrrrr}
     & \multicolumn{1}{c@{\qquad}}{\textbf{Numerical}} & \multicolumn{4}{c}{\textbf{Symbolic}}\\
    & Mean   
                      & Mean   & Median & StdDev & Maximum \\
    026 -- Original 	& 2.4~    
                      & 0.568  & 0.530  & 0.107 & 0.905   \\
    026 -- Reduced  	& 0.85   
                      & 0.053  & 0.047  & 0.036 & 0.343   \\
    028 -- Original 	& 16.57  
                      & 42.430 & 40.529 & 8.632 & 84.116 \\
    028 -- Reduced  	& $\bot$      
                      & 0.485  & 0.468  & 0.119 & 0.796       
  \end{tabular}
\end{table}

\subsection{Going Further}
\label{SEC:3d}

\begin{figure}[p]
  \centering
  \includegraphics[width=0.44\textwidth]{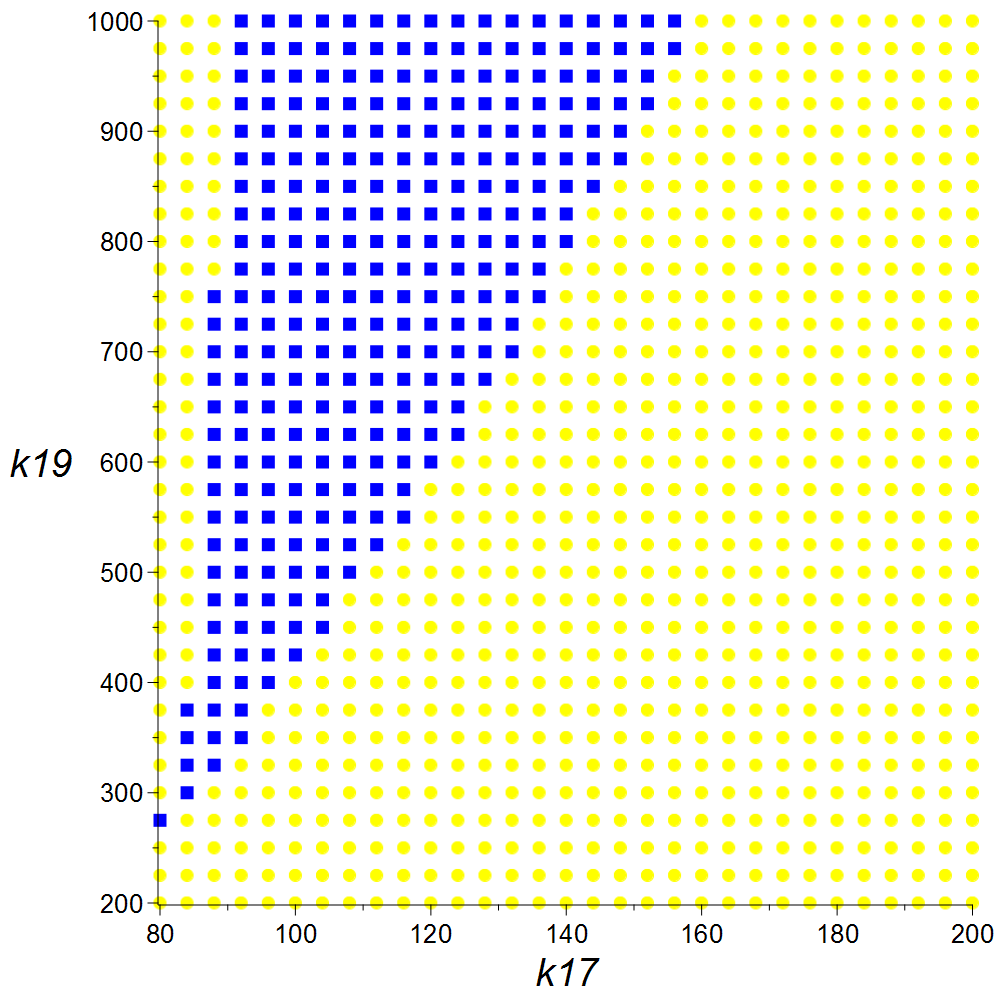}
  \includegraphics[width=0.44\textwidth]{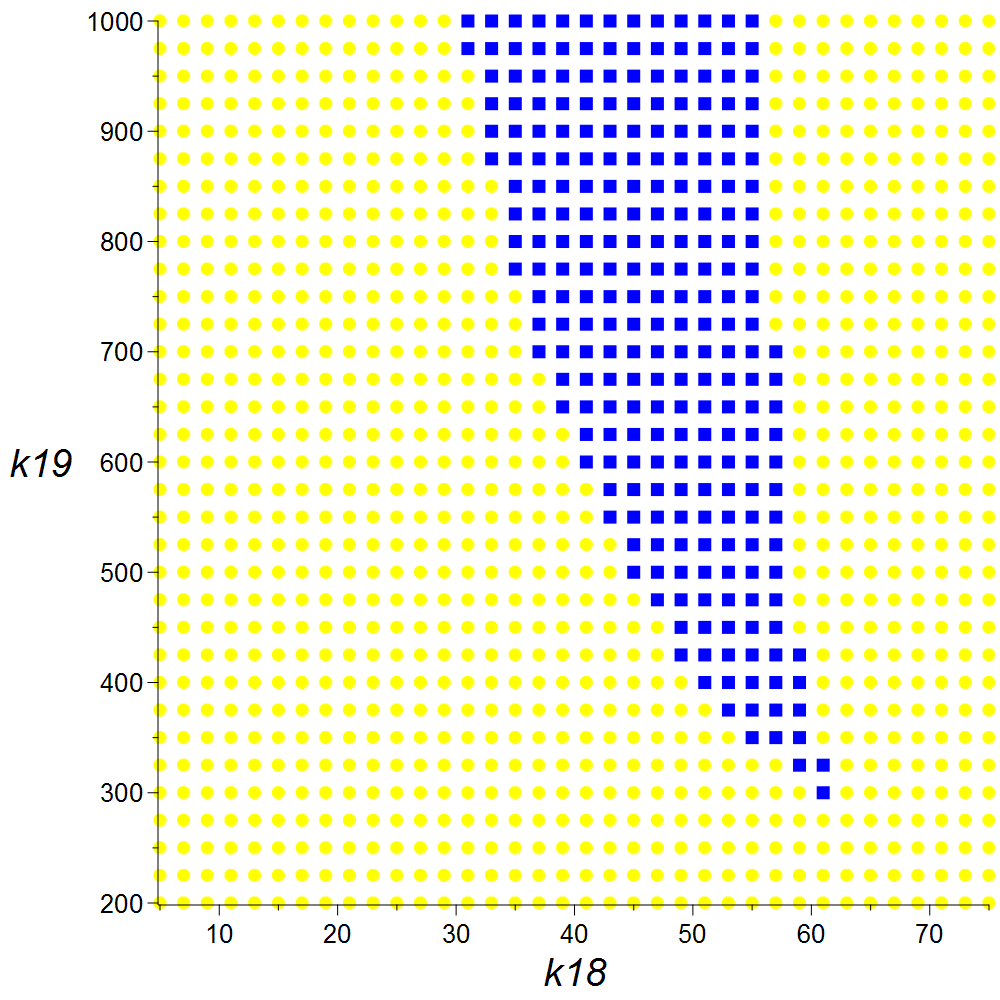}
  \caption{As Fig.~\ref{FIG:Maple-Sys26} but with a higher sampling rate\label{FIG:Sys26Detailed}}
\end{figure}

Of course, we could increase the sampling density to get an improved idea of the bistability region, as in Fig.~\ref{FIG:Sys26Detailed} and Fig.~\ref{FIG:Sys28Detailed}.  
However, a greater understanding comes with 3D sampling.  We have performed this using the symbolic approach described above, at a linear cost proportional to the increased 
number of sample points.  This was completed for \btwsix: the region in question is bounded to both sides in the $k_{17}$ and $k_{18}$ directions but extends infinitely 
above in $k_{19}$.  With the $k_{19}$ range bound at 1000 the region is bounded by extending $k_{17}$ to 800 and $k_{18}$ to 600.  
For obtaining exact bounds (in one parameter) see \cite{Bradford2017}.  

\begin{figure}[p]
  \centering
  \includegraphics[height=0.25\textheight, width=0.45\textwidth]{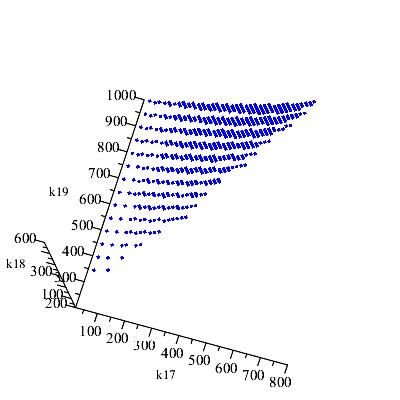}
  \includegraphics[height=0.25\textheight, width=0.45\textwidth]{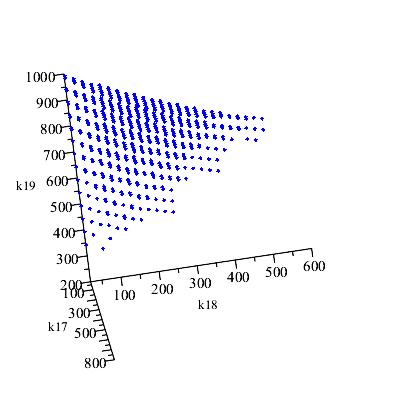}
  \caption{3D Maple Point Plot produced grid sampling on \btwsix (see Sect.~\ref{SEC:3d})\label{FIG:3dPointPlot}}
\end{figure}

\begin{figure}[p]
  \centering
  \includegraphics[height=0.25\textheight, width=0.45\textwidth]{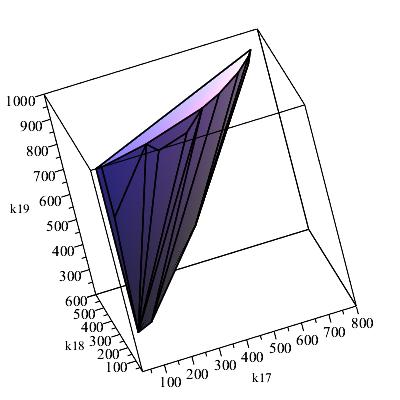}
  \includegraphics[height=0.25\textheight, width=0.45\textwidth]{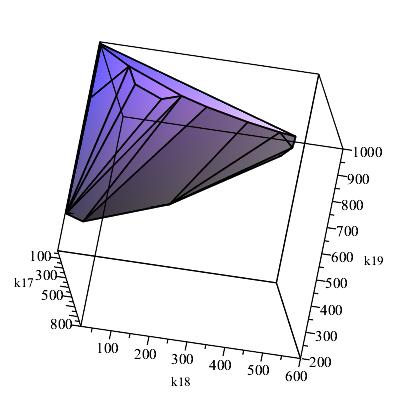}
  \caption{Convex Hull of the bistable points in Fig.~\ref{FIG:3dPointPlot}\label{FIG:3dConvexHull}}
\end{figure}

Sampling in 20 seconds for $k_{17}$ and $k_{18}$ and 50 seconds for $k_{19}$
produced a Maple point plot of 20400 in 18 minutes. Figure \ref{FIG:3dPointPlot}
shows 2D captures of the 3D bistable points and Fig.~\ref{FIG:3dConvexHull} the
convex hull of these, produced using the convex
package\footnote{\url{http://www.math.uwo.ca/~mfranz/convex/}}.
We note the lens shape seen in the orientation in the left
plots is comparable with the image
 in the original paper of Markevich et al.~\cite[Fig.~S7]{Markevich2004}.

\section{Conclusion and Future Work}

We described a new graph theoretical symbolic preprocessing method to reduce
problems from the MAPK network. We experimented with two systems and found the
reduction offered computation savings to both numerical and symbolic approaches
for the determination of multistationarity regions of parameter space. In
addition, the reduction avoided instability from rounding errors in the
numerical approach \pagebreak to one system, but uncovered major problems in that approach
for the other. An interesting side result is that, at least for the smaller
system, the symbolic approach can compete with and even outperform the numerical
one, demonstrating how far such methods have progressed in recent years.

In future work we intend to combine the results of the present paper and our
recent publication \cite{Bradford2017} to generate symbolic descriptions of the
bistability region beyond the 1-parameter case. Other possible routes to achieve
this is to consider the effect of the various degrees of freedom with the
algorithms used. For example, we have a free choice of variable ordering:
\btwsix has 11 variables corresponding to 39\,916\,800 possible orderings while
\btweight has 16 variables corresponding to
more than $10^{13}$ orderings. Heuristics exist to help with this choice
\cite{DSS:04a} and machine learning may be applicable \cite{HEWDPB14}. Also,
since MAPK problems contain many equational constraints an approach as described
in \cite{EBD15} may be applicable when higher dimensional CADs are needed.


\section {Acknowledgements}
D.~Grigoriev is grateful to the grant RSF 16-11-10075. H.~Errami, O.~Radulescu, and
A.~Weber thank the French-German Procope-DAAD program for partial support of
this research. M.~England and T.~Sturm are grateful to EU H2020-FETOPEN-2015-CSA
712689 SC\textsuperscript{2}.

\vspace*{0.1in}

\textbf{Research Data Statement:} Data supporting the research in this paper is available from \href{http://doi.org/10.5281/zenodo.807678}{doi:10.5281/zenodo.807678}.

\bibliographystyle{splncs_srt}
\bibliography{casc2017}
\end{document}